\documentstyle[12pt]{article}

\newcommand{\Eq}[1]{Eq.~(\ref{#1})}
\newcommand{\Eqs}[2]{Eq.~(\ref{#1}) and (\ref{#2})}
\newcommand{\Ref}[1]{Ref.~\cite{#1}}
\newcommand{\beq}[1]{\begin{equation}\label{#1}}
\newcommand{\eeq}{\end{equation}}
\newcommand{\bdm}{\begin{displaymath}}
\newcommand{\edm}{\end{displaymath}}
\newcommand{\beqa}[1]{\begin{eqnarray}\label{#1}}
\newcommand{\eeqa}{\end{eqnarray}}
\newcommand{\bdma}{\begin{eqnarray*}}
\newcommand{\edma}{\end{eqnarray*}}

	\newcommand{\prd}[3]{Phys. Rev. D{\bf #1}, #2 (#3)}
	\newcommand{\physrep}[3]{Phys. Rep. {\bf #1}, #2 (#3)}
	\newcommand{\plb}[3]{Phys. Lett. {\bf B#1}, #2 (#3)}
	\newcommand{\npb}[3]{Nucl. Phys. {\bf B#1}, #2 (#3)}
	\newcommand{\prl}[3]{Phys. Rev. Lett. {\bf #1}, #2 (#3)}
	\newcommand{\rmp}[3]{Rev. Mod. Phys. {\bf #1}, #2 (#3)}
	\newcommand{\ibid}[3]{{\em ibid.} {\bf #1}, #2 (#3)}
	\newcommand{\astropj}[3]{Ap. J. {\bf #1}, #2 (#3)}

	\newcommand{\sovphys}[3]{Sov.  J.  Nucl. Phys. {\bf #1}, #2 (#3)}

\newcommand{\aslash}[1]{{\rlap/#1}}
\newcommand{\splash}[1]{{#1\mkern -9.0mu /}}		

\date{November 1994\\LTP-044-UPR}
\title{Field-theoretic treatment of mixed neutrinos in a
neutrino and matter background}
\author{J. C. D'Olivo\thanks{
Partially supported by
Grant No. DGAPA-IN100691}\\
	Instituto de Ciencias Nucleares\\
	Universidad Nacional Aut\'{o}noma de M\'{e}xico\\
	Apartado Postal 70-543, 04510 M\'{e}xico, D.F., M\'{e}xico
	\and Jos\'{e} F. Nieves\thanks{
Partially supported by the
US National Science Foundation Grant PHY-9320692}\\
	Laboratory of Theoretical Physics\\
	Department of Physics, P. O. Box 23343\\
	University of Puerto Rico\\
R\'{\i}o Piedras, Puerto Rico 00931-3343
}

\begin{document}

\maketitle

\begin{abstract}

We use the method  of finite temperature field theory to examine
the propagation of mixed neutrinos through dense media,
putting the emphasis in those situations
in which the neutrinos themselves are in
the background.
The evolution equation for the flavor amplitudes is deduced, and
the expressions for the corresponding hamiltonian
matrix are given explicitly.  We find that, in order to include
the nonlinear effects
due to the $\nu$-$\nu$ interactions, the neutrino
propagator that must be used
in the calculation of the neutrino self-energy diagrams
that contain neutrinos in the internal lines is
the propagator for the neutrino modes in the medium
instead of the thermal free-field propagator.
We also show how the absorptive contributions are included
in terms of a non-hermitian part of the hamiltonian matrix,
which we indicate how it is calculated.
Our treatment provides a consistent generalization of a
method that has been successfully
applied to the study of neutrino oscillations in matter.

\end{abstract}
\section{Introduction and Summary}\label{sec:introduction}

Finite Temperature Field Theory (FTFT)
provides a very useful framework  to study
the behavior of elementary particles, including neutrinos,
in the presence
of a thermal background\cite{FTFT}.
A quantity of fundamental interest in this formalism
is the  self-energy of a particle,
from which the dispersion relation of the
modes that propagate
in the medium  are obtained.
In the case of neutrinos,
the density and temperature dependent contributions
to the self-energy can be expressed as an effective potential
for each neutrino flavor that,
added to the vacuum kinetic energy,  yield
the Hamiltonian matrix  that  governs
the neutrino oscillations in matter\cite{Wolfenstein,MSmirnov}.
In normal matter (electrons and nucleons) the universality of the
neutral-current interactions of the neutrinos implies that their
contribution to the effective potential is the same
for all flavors and hence irrelevant as far as
oscillations are concerned.
However, in environments like the early universe or the core of a
supernova, where the neutrinos represent an appreciable fraction
of the total density, the contributions to the potential energy
arising from the $\nu$-$\nu$ scattering are not in general
proportional to the unit matrix
and should be included in the analysis of the
resonant flavor transformations.

The calculation of the contribution
from the background neutrinos
to the effective potential presents a new challenge because
now the background depends itself on the flavor amplitudes,
whose time evolution is in turn
determined by the oscillation mechanism.
Thus the problem becomes a non-linear one that must
be solved in a self-consistent  manner.
Several studies of  the astrophysical and cosmological
implications of neutrino oscillations
under these conditions
have been carried out during the last
years\cite{earlierlit,cline,densitymatrix}.
These works
left out the off-diagonal
contributions of the $\nu-\nu$ scattering to the flavor energy matrix.
The  existence of these extra terms was first recognized by
Pantaleone\cite{pantaleone} and later on have been included in
certain numerical analysis of neutrinos oscillations\cite{samuel}.

Most of the extensive studies
of neutrino oscillations in matter\cite{mswreviews} and the
MSW solution to the solar neutrino problem\cite{solarnus} are
based on the Wolfenstein equation for
the flavor amplitudes of a
single-particle wave function.
On the other hand,
according to the prevailing attitude in the literature on the subject,
a consistent description
of neutrino oscillations that includes the effects of their mutual
interactions and the noncoherent interactions
with the background
is possible only within the density matrix
formalism\cite{densitymatrix,densitymatrix2,siglraffelt,mckellar}.
The works cited in \Ref{siglraffelt,mckellar} give a general treatment
of relativistic mixed neutrinos along this line which include
the neutrino self-interactions, and have been recently applied
to examine the matter-enhanced flavor transformations
in supernovae\cite{fuller,pantaleone2}.

While it is widely accepted that
FTFT is an efficient method to compute the background effects
that are incorporated as a potential energy
in the Wolfenstein equation,
a description based on this approach
when neutrinos compose the background is lacking
or incomplete.
The present work fills this gap.
By working within the formalism of the FTFT,
we derive the  evolution equation for the
flavor amplitudes of a neutrino that propagates in a medium
containing  several species of massive neutrinos.
Explicit formulas for the energy matrix are given
in a familiar form that make them amenable
for application to concrete situations.
In order to bring out the salient features without
introducing unnecessary complications, we work
in detail the case of two Dirac neutrinos in the Standard Model,
but our approach makes evident
the path to treat more general cases, such as including
additional families and Majorana neutrinos.

One of our objectives is to show that FTFT, which is simpler to use
than other approaches, can be consistently applied
to examine  the type of problem  we have referred to.
The procedure is to calculate the self-energy matrix
for mixed neutrinos, which leads to the effective
Dirac equation  obeyed by the modes that
propagate in the medium.
We obtain the Hamiltonian matrix for the flavor evolution equation
directly from the condition that determines
the energy-momentum relation for the neutrino modes.
A particular new feature that arises
when neutrinos appear in the background
is that the
neutrino densities are modulated by the oscillation
mechanism itself, but on the other hand the
standard FTFT propagators do not take this into account.
Therefore an essential part of the problem is
to find the appropriate neutrino propagator to be used
in the self-energy diagrams that contain neutrinos in the
internal lines.  This aspect of the problem
is considered in Section~\ref{sec:nubackground},
where we show that this
is not the standard free-particle thermal propagator,
but rather the one for the effective field
constructed out of the one-particle wave functions
of the propagating modes.  The use of this
propagator is a key ingredient in our treatment.
We also show how to include the
damping effects by means of a non-hermitian part
in the Hamiltonian\cite{cline},
which is determined in our scheme from
the absorptive part of the self-energy.  In
Section~\ref{sec:abseffects}
we prove that the non-hermitian
part of the Hamiltonian can be expressed
in terms of the total rates for the emission and
absorption processes of the mixed neutrinos by the background.
There we also discuss some subtleties that arise
if these rates are determined in the traditional
way, in terms of probability amplitudes,
when neutrinos compose the background and we explain
how they are overcome in our treatment.

\section{Normal Matter}\label{sec:normalmatter}
The dispersion relations and wave functions of the neutrino
modes that propagate through a medium are determined from
the linear part of the effective field equation.
In momentum space it takes the form
\beq{fieldeq}
\left(\splash k - m - \Sigma_{eff}\right)\psi  = 0\, ,
\eeq
where $k_{\mu}$ is the momentum vector and
$m$ is the diagonal matrix whose elements are the
neutrino masses in vacuum.
In the basis of the mass eigenstates $\nu_i $, the background contributions
to the neutrino self-energy
$\Sigma_{eff}$
is a non-diagonal matrix
which can be written as\cite{Weldon:fermions}
\beq{Sigma}
(\Sigma_{eff})_{ij} = (a_{Lij}\aslash k + b_{Lij}\aslash u + c_{Lij})L +
(a_{Rij}\aslash k + b_{Rij}\aslash u + c_{Rij})R \,,
\eeq
where $u^\mu$ is the velocity four-vector of the medium and
$L,R = (1\mp\gamma^5)/2$.
In the rest frame of the medium
\bdm
u^{\mu} = \left(1,\vec{0}\right) \,,
\edm
and the components of $k^{\mu}$ are given by
\bdm
k^{\mu} = \left(\omega,\vec{\kappa}\right)  \,.
\edm

In general, the elements of the matrices $a_{L,R}$, $b_{L,R}$ and $c_{L,R}$
in \Eq{Sigma}
depend of $\omega$ and $\vec\kappa$, but  in a isotropic
medium they are functions only of the Lorentz scalars
\beqa{variables}
\omega & = & k\cdot u\,, \nonumber \\
\kappa & = & \sqrt{{\omega}^{2} - k^{2}}\,,
\eeqa
with $\kappa = \left|\vec{\kappa}\right|$  .
In what follows, the arguments indicating the dependence on these variables
will be generally omitted,  but they will be explicitly
included  when needed for clarity.

The background  is taken to consist
of nucleons,  electrons and neutrinos,
and their respective antiparticles. We assume that the
temperature is low enough so that
the contributions of the heavier charged leptons can be neglected.
The relevant piece of the weak interaction Lagrangian is
\beqa{lagrangian}
L_{\mbox{int}} & = &
-\frac{g}{2\cos\theta_W}Z^\mu
\left[\sum_i \overline\nu_{Li}\gamma_\mu \nu_{Li}\right.
\mbox{}+ \left.\sum_{f=e,n,p}
\overline f\gamma_\mu(X_f + Y_f\gamma^5) f\right]\nonumber\\
& & \mbox{}
- \frac{g}{\sqrt{2}}W^\mu\sum_i U_{e i}\overline e_L
\gamma_\mu\nu_{Li}\,,
\eeqa
where the index $i$ runs over all the neutrino species.
The $\nu_{Li}$ are the left-handed components of the neutrino fields
with a definite mass, and $U$
is the unitary matrix that relates them with the neutrino flavor
fields $\nu_{L\alpha}=\sum_i U_{\alpha i} \nu_{Li}$
($\alpha = e, \mu, \tau)$.  The coefficients
$X_f$ and $Y_f$ are
\beqa{exycouplings}
X_e & = & -\frac{1}{2} + 2\sin^2\theta_W\,,\nonumber\\
Y_e & = & \frac{1}{2}\,,
\eeqa
for the electron and
\beqa{nxycouplings}
X_p & = & \frac{1}{2} - 2\sin^2\theta_W\,,\nonumber\\
Y_n & = & -X_n = -Y_p = \frac{1}{2}\,,
\eeqa
for the nucleons.  In addition,
\beqa{GFermi}
m_Z\cos\theta_W & = & m_W\,,\nonumber\\
\frac{g^2}{4m_W^2} & = & \sqrt{2}G_F\,.
\eeqa

The calculation of the self-energy
proceeds as in the vacuum, but with the free
propagators for the internal lines replaced by their
thermal generalizations.  Since we are assuming
that there are no $W$ and $Z$ bosons in the medium,
their propagators are the same as in the vacuum.

Let us first discuss the simplest case of a
homogeneous background
that does not contain neutrinos.
For Dirac neutrinos propagating in such a
background, the result of the
one-loop calculation of the self-energy,
up to terms of order $g^2/m_W^2$, is\cite{DNT}
\beqa{coefficients}
a_{L,Rij} & = & c_{L,Rij}\  = \  0\,,\nonumber\\
b_{Rij} & = & 0\,,\nonumber\\
b_{Lij} & = &  b_{ij}\  =\   \sqrt{2}G_F\left[U^\ast_{ei}U_{ej}(n_e -
n_{\overline e}) +
\delta_{ij}Q_Z\right]\,,
\eeqa
where
\beq{QZ}
Q_Z = \sum_{f = e,n,p}{X_f(n_f - n_{\overline f})}\,,
\eeq
is the average $Z$-charge of the medium.  In these formulas,
$n_f$ and $n_{\overline f}$ denote
the total number
densities of the particles (electrons, neutrons and protons)
and the antiparticles, respectively.

In terms of the left- and right-handed components of $\psi$
\Eq{fieldeq} becomes
\beqa{eq6.4}
\splash A_L\psi_L - m\psi_R & = 0\,,\nonumber\\
\splash A_R\psi_R - m\psi_L & = 0\,,
\eeqa
where
\beqa{eq6.5}
A_{\mu L}  & = & k_\mu - b u_\mu\,,\nonumber\\
A_{\mu R}  & = & k_\mu\,.
\eeqa
In writing these equations we have used the particular
results given in \Eq{coefficients}.
In a more general case $A_{\mu L,R}  = (1 - a_{L,R})k_\mu - b_{L,R}u_\mu$
and the matrix $m$ in \Eq{eq6.4} has to be replaced by $m + c_{R,L}$ in
the first and in the second equation, respectively.

We use the
Weyl representation of the gamma matrices and put
\beqa{eq6.6}
\psi_R & = & \left( \begin{array}{c}\xi\\0\end{array}\right)\,,\nonumber\\
\psi_L & = & \left( \begin{array}{c}0\\\eta\end{array}\right)\,.
\eeqa
The equations to be solved then become
\beqa{eq6.7}
\left(A_L^0 + \vec{\sigma}\cdot\vec A_L\right)\eta - m\xi & = & 0\,,\nonumber\\
\left(A_R^0 - \vec{\sigma}\cdot\vec A_R\right)\xi - m\eta & = & 0\,.
\eeqa
Since  $\vec{A_{L}}$ and $\vec{A_{R}}$ are both proportional to
$\vec{\kappa}$, these equations can have non-trivial solutions only
if $\xi$ and $\eta$ are proportional
to the same spinor $\phi_\lambda$ with definite helicity,
defined by
\beq{eq6.8}
\left(\vec\sigma\cdot\hat k\right)\phi_\lambda = \lambda\phi_\lambda\,,
\eeq
with $\lambda = \pm 1$. Therefore, we write
\beqa{eq6.9}
\xi & = & y\phi_\lambda\,,\nonumber\\
\eta & = & x\phi_\lambda\,,
\eeqa
where $y$ and $x$ are spinors in the basis of the $\nu_i$.
For a given helicity $\lambda$, the equations for $y$ and $x$
that follow from \Eq{eq6.7} are
\beqa{eq6.10}
A_L^{(\lambda)}x - my & = & 0\,,\nonumber\\
A_R^{(-\lambda)}y - mx & = & 0\,,
\eeqa
where
\beqa{eq6.11}
A_{L}^{(\lambda)} & = & \omega + \lambda\kappa - b\,,\nonumber\\
A_{R}^{(\lambda)} & = & \omega + \lambda\kappa\,.
\eeqa
Using the second equation in (\ref{eq6.10}) to eliminate $y$ from
the first one, the equations for $y$ and $x$ become
\beqa{eq.r1}
\left[A_R^{(-\lambda)}m^{-1}A_L^{(\lambda)} -m\right ]x & = & 0\,,\nonumber\\
y & = & \frac{1}{A_R^{(-\lambda)}}\ mx\,,
\eeqa
which have solutions only if
\beq{Det}
\mbox{Det}\left(A_R^{(-\lambda)} m^{-1}A_L^{(\lambda)} - m\right) = 0.
\eeq

The roots of this equation determine the dispersion
relations of the propagating modes.
The complete characterization of the different solutions
requires, in addition,
the determination of the wave function for each mode.
For these purposes
it is instructive to notice that
for one (unmixed) Dirac neutrino, \Eq{Det} reduces to
\beqa{dispreleqleft}
(\omega - \kappa - b)(\omega + \kappa) & = & m^2\,,\\
\label{dispreleqright}
(\omega + \kappa - b)(\omega - \kappa) & = & m^2\,,
\eeqa
where the expression for $b$ is obtained from \Eq{coefficients}
by making obvious simplifications.
For example, for an electron neutrino in normal matter,
$b = \sqrt{2}G_F(n_e +
\sum_{f }{X_fn_f})$.
Each one of Eqs.~(\ref{dispreleqleft}) and
(\ref{dispreleqright}) yields two solutions: one with
a positive real part of $\omega$ and another with a negative real part.
The four solutions correspond to the positive and negative
helicity states
of the neutrino and the antineutrino.
In particular, the dispersion relation for the left-handed
neutrino corresponds to the positive solution of \Eq{dispreleqleft}:

\beq{disprelsol}
\omega = \left[(\kappa + \frac{b}{2})^2 + m^2\right]^{1/2} +
\frac{b}{2}\,.
\eeq
For future purposes it is useful to notice here that,
for relativistic
neutrinos, the effect on the dispersion relations of a non-zero value
of the coefficients $a_L$ and $a_R$ is negligible.  To see this notice,
for example, that if $a_L\not = 0$ \Eq{dispreleqleft} is modified to
\beq{dispreleqleft2}
[(1 - a_L)(\omega - \kappa) - b](\omega + \kappa) = m^2\,.
\eeq
For $\omega\approx\kappa$, the corrections to the solution
given in \Eq{disprelsol} are either of higher order
in $g^2$ or of order
$m^2g^2/\kappa$ and we ignore them.

Based on these remarks, let us now return to \Eq{Det}
and look for solutions
with $\lambda = -1$.  As already mentioned, for
clarity we consider
the case of mixing
between only two Dirac neutrinos.  The
condition that determines the energy-momentum relation is given explicitly
by
\beqa{eq6.12}
\left[\omega^2 - \kappa^2 - (\omega + \kappa)b_{11} - m_1^2\right]
\left[\omega^2 - \kappa^2 - (\omega + \kappa)b_{22} - m_2^2\right]
\nonumber\\
\mbox{} - (\omega + \kappa)^2b_{12}b_{21} = 0\,,
\eeqa
where
\beqa{eq6.13}
b_{11} & = & b_e\cos^2\theta + \sqrt{2}G_F Q_Z\nonumber\\
b_{22} & = & b_e\sin^2\theta + \sqrt{2}G_F Q_Z\nonumber\\
b_{12} & = & b_{21} = b_e\sin\theta\cos\theta\,.
\eeqa
Here $\theta$ is the vacuum mixing angle and
\beq{be}
b_e = \sqrt{2}G_F(n_e - \overline n_e)\,.
\eeq
\Eq{eq6.12} has in general two positive and two negative solutions,
the latter of which correspond to the antineutrinos.
We seek approximate solutions to this equation in the
relativistic limit, which is the adequate regime for
neutrino oscillations.  We do it first for neutrinos (positive solutions), and
afterwards
apply the same considerations to the antineutrinos.
Dividing \Eq{eq6.12} by $(\omega + \kappa)^2$ and using
the approximation
\beq{eq.r24}
\frac{m_i^2}{\omega + \kappa}\simeq\frac{m_i^2}{2\kappa}\,,
\eeq
which is valid for $\kappa\gg m_{1,2}, b_{ij}$,
\Eq{eq6.12} reduces to
\beq{eq6.13.1}
\left(\omega - \kappa - b_{11} - \frac{m_1^2}{2\kappa}\right)
\left(\omega - \kappa - b_{22} - \frac{m_2^2}{2\kappa}\right)
\nonumber\\
\mbox{} - b_{12}b_{21} = 0\,.
\eeq
with solutions
\beq{eq6.16}
\omega_{1,2} = \kappa +  \frac{m_1^2 + m_2^2}{4\kappa}
+ \frac{b_{11} + b_{22}}{2}
\mp\frac{1}{2} \left[{\left(b_{11} - b_{22} + \frac{m_1^2 -
m_2^2}{2\kappa}\right)^2
+ 4b_{12}b_{21}}\right]^{1/2}\,.
\eeq
With each dispersion relation $\omega_{1,2}$ there is associated a
two-component vector $x = e_{1,2}$, which is obtained by solving
\beq{eq.r2}
\left.\left[A_R^{(+)}m^{-1}A_L^{(-)} - m\right ]\right|_{\omega =
\omega_{1,2}}e_{1,2} = 0\,.
\eeq
The corresponding $y$-vector is given by
\beq{eq.r3}
y_{1,2} = \left.\frac{m}{A_R^{(+)}}\right|_{\omega = \omega_{1,2}}e_{1,2}\,,
\eeq
and together with
Eq. (\ref{eq.r2}), it  determine the wave function
of the propagating mode.
Clearly, for relativistic neutrinos,
$y_{1,2}\approx(m/2\kappa)e_{1,2}\,$.

Using the same approximation
of (\ref{eq.r24}), Eq. (\ref{eq.r2}) can be recast in the form
\beq{eq.r4}
He_i = \omega_i e_i\,,
\eeq
where
\beq{Hamgeneral}
H = \kappa + \frac{m^2}{2\kappa}  +  b(\kappa,\vec\kappa)\,.
\eeq
For the two-flavor mixing case we are considering,
\beq{Ham}
H  =  \kappa + \sqrt{2}G_FQ_Z + \frac{1}{2\kappa}\left(\begin{array}{cc}
m_{1}^2 & 0\\
0 & m_{2}^2
\end{array}\right)  +
U^\dagger\left(\begin{array}{cc}
b_e & 0\\
0 & 0
\end{array}\right)U\,.
\eeq
with
\beq{Udos}
U =
\left(\begin{array}{cc}
\cos\theta& \sin\theta\\
-\sin\theta &\cos\theta
\end{array}\right)\,.
\eeq
The corresponding Hamiltonian in the flavor basis is
\beq{Hamflavor}
H_{flav} = UHU^\dagger\,,
\eeq
and after subtracting an irrelevant term proportional
to the unit matrix, it can be written in the form
\beqa{Hamflavusual}
H_{flav} =
\frac{1}{2}\left(\begin{array}{cc}
-\Delta_0\cos 2\theta  + b_e &  \Delta_0\sin2\theta\\
 \Delta_0\sin2\theta & \Delta_0\cos 2\theta  - b_e
\end{array}\right)\,,
\eeqa
where
\beq{Delta0}
\Delta_0 = \frac{m_2^2 - m_1^2}{2\kappa}\,.
\eeq
This is the energy matrix
traditionally used in the studies of
neutrino oscillations inside the Sun.  Its eigenvalues are
simply $\mp\frac{1}{2} (\omega_2 - \omega_1)$, with
\beq{omegapm}
\omega_2 - \omega_1 =
\sqrt{\left(\Delta_0\cos 2\theta  - b_e \right )^2
+ \Delta_0^2\sin^2 2\theta}\,.
\eeq

The explicit solutions of Eq.~(\ref{eq.r2}) in the flavor basis
are readily found to be
\beqa{eq.r5}
e_1 & = & \left( \begin{array}{c}\cos\theta_m\\-\sin\theta_m\end{array}\right),
 \nonumber\\
e_2 & = &\left( \begin{array}{c}\sin\theta_m\\\cos\theta_m\end{array}\right
)\,,
\eeqa
where $\theta_m$ is the mixing angle in the medium given by
\beq{eq.r5.1}
\sin2\theta_m =\frac{\Delta_0\sin2\theta}{\sqrt{\left(\Delta_0\cos 2\theta  -
b_e\right )^2
+ \Delta_0^2\sin^2 2\theta}}\,.
\eeq

Written as in \Eq{eq.r4}, the dispersion relations $\omega_{1,2}$
are simply the eigenvalues of the Hamiltonian $H$ with the $e_i$
being the corresponding eigenvectors.  The physical interpretation
that emerges is the following.
The Dirac wave function for a relativistic left-handed neutrino with
momentum $\vec\kappa$ that propagates through a medium with a definite
dispersion relation $\omega_i$ is
\beqa{eq.r7}
\psi (x) & \simeq & \psi_L (x)\nonumber\\
& = &  e^{i\vec \kappa\cdot \vec x}
\left( \begin{array}{c}0\\\phi_{-}\end{array}\right)
e^{-i\omega_it}e_i\,,
\eeqa
where we neglect $\psi_R$, which is of order $m/2\kappa$
as compared to the left-handed component.
If the neutrino is not initially in a state corresponding to one
of the  eigenmodes, then the wave function is
\beq{eq.r8}
\psi_L = e^{i\vec\kappa\cdot \vec x}
\left( \begin{array}{c}0\\\phi_{-}\end{array}\right)\chi(t)\,,
\eeq
where $\chi(t)$ is the general solution of
\beq{eq.r9}
i\frac{d\chi}{dt} = H\chi\,,
\eeq
that satisfies the
specified boundary condition for $\chi(0)$,
and can be written as
\beq{eq.r10}
\chi(t) = \sum_{i=1,2}(e_i^\dagger \chi(0))e^{-i\omega_it} e_i \,.
\eeq
Since neutrinos are produced in states of definite flavor,
in neutrino oscillations problems the initial conditions are commonly
specified by giving the flavor components of the
initial state.
Thus, if we denote
\beq{eq.r11}
\chi(0) = \left( \begin{array}{c}a_e\\a_\mu\end{array}\right)\,,
\eeq
in the flavor basis, then at a later time $t$
\beq{eq.r12}
\chi(t) = \chi^{(e)}(t)a_e + \chi^{(\mu)}(t)a_\mu\,,
\eeq
where $\chi^{(e,\mu)}(t)$ are the solutions to \Eq{eq.r9} with the
initial conditions
\beqa{eq.r13}
\chi^{(e)}(0) & = & \left( \begin{array}{c}1\\0\end{array}\right)\,,\nonumber\\
\chi^{(\mu)}(0) & = & \left( \begin{array}{c}0\\1\end{array}\right)\,.
\eeqa
Physically, $a_{e,\mu}$ is the amplitude for finding the
neutrino initially in the state corresponding to
$|\nu_{e,\mu}\rangle$.  If we write
\beq{eq.r14}
\chi(t) = \left( \begin{array}{c}\chi_e(t)\\\chi_\mu(t)\end{array}\right)
\eeq
then
\bdm
\begin{array}{cc}
\chi_{\alpha}(t) =  a_e \chi_{\alpha}^{(e)}(t) +
a_{\mu} \chi_{\alpha}^{(\mu)}(t)\,, & \mbox{($\alpha = e,\mu$)}
\end{array}
\edm
is the amplitude for finding
the neutrino in the corresponding  flavor state
at subsequent times, with a similar interpretation
for the components
of $\chi(t)$ in other basis.

Associated with the wave functions
there is a  quantum field
\beq{eq.r15}
\Psi_L^{(\nu)}(x) = \int\frac{d^3\kappa}{(2\pi)^3}e^{i\vec \kappa\cdot \vec x}
\left(
\begin{array}{c}0\\\phi_{-}\end{array}\right)(\chi^{(e)}(t)a_e(\vec\kappa) +
\chi^{(\mu)}(t)a_\mu(\vec\kappa))\,,
\eeq
where $\chi^{(e,\mu)}$ are the functions defined above,
while the $a_{e,\mu}(\vec\kappa)$
are now interpreted as the
annihilation operators of left-handed neutrinos of definite
flavor with momentum $\vec\kappa$.

Up to now we have restricted ourselves to the positive
frequency solutions of \Eq{Det} with $\lambda = -1$.
In order to include the right-handed antineutrinos we
need to consider
the negative solutions with  $\lambda = +1$.
Writing the negative solutions as $\omega = -\overline\omega$,
then $\overline\omega_{1,2}$ are given by the same formulas as
in \Eq{eq6.16} but with the sign of the coefficients $b_{ij}$ reversed:
\beq{eq.r16}
\overline\omega_{1,2} = \kappa +  \frac{m_1^2 + m_2^2}{4\kappa}
- \frac{b_{11} + b_{22}}{2}
\mp\frac{1}{2} \left[{\left(b_{22} - b_{11} + \frac{m_1^2 -
m_2^2}{2\kappa}\right)^2
+ 4b_{12}b_{21}}\right]^{1/2}.
\eeq
The equations corresponding to \Eq{eq.r2}
for the two-component vectors $x = \overline e_{1,2}$ are
\beq{eq.r17}
\left.\left[A_R^{(-)}m^{-1}A_L^{(+)} - m\right ]\right|_{\omega =
-\overline\omega_{1,2}}
\overline e_{1,2} = 0\,.
\eeq
Similarly to \Eq{eq.r4}, they can be recast in the form
\beq{eq.r18}
H_{-}\overline e_i = \overline\omega_i \overline e_i\,,
\eeq
where
\beq{Hamneggeneral}
H_{-} = \kappa + \frac{m^2}{2\kappa} - b(-\kappa,\vec\kappa)\,.
\eeq
In particular, $H_{-}$ differs from \Eq{Ham} only in the sign of the $Q_Z$ and
$b_e$ terms.
In analogy with the positive frequency field $\Psi^{(\nu)}_L$, we built a
negative frequency field
\beq{eq.r19}
\Psi_L^{(\overline\nu)}(x) = \int\frac{d^3\kappa}{(2\pi)^3}e^{-i\vec
\kappa\cdot \vec x}
\left( \begin{array}{c}0\\\
i\sigma_2\phi_{+}^\ast\end{array}\right)(\chi^{(\overline e)}(t)a_{\overline
e}(\vec\kappa) + \chi^{(\overline \mu)}(t)a_{\overline
\mu}(\vec\kappa))^\ast\,,
\eeq
where the $a_{\overline e, \overline \mu}(\vec\kappa)$
represent the annihilation operators of (right-handed) antineutrinos of
definite
flavor, with
momentum $\vec\kappa$.  The
$\chi^{(\overline e,\overline \mu)}$ are solutions of
\beq{eq.r20}
i\frac{d\chi}{dt} = \overline H\chi\,,
\eeq
that satisfy
\beqa{eq.r21}
\chi^{(\overline e)}(0) & = & \left(
\begin{array}{c}1\\0\end{array}\right)\,,\nonumber\\
\chi^{(\overline\mu)}(0) & = & \left( \begin{array}{c}0\\1\end{array}\right)\,,
\eeqa
with
\beqa{overlineHam}
\overline H  & \equiv & H^{\ast}_{-}(-\vec\kappa)\,,\nonumber\\
& = & \kappa + \frac{m^2}{2\kappa} -
b_L^\ast(-\kappa,-\vec\kappa)\,,\nonumber\\
& = &   \kappa - \sqrt{2}G_FQ_Z + \frac{1}{2\kappa}\left(\begin{array}{cc}
m_{1}^2 & 0\\
0 & m_{2}^2
\end{array}\right)  -
U^T\left(\begin{array}{cc}
b_e & 0\\
0 & 0
\end{array}\right)U^\ast\,.
\eeqa
The corresponding matrix in the flavor basis is
\beqa{overlineHamflavor}
\overline H_{flav}  & = &  U^\ast\overline HU^T\nonumber\\
& = &\frac{1}{2}\left(\begin{array}{cc}
-\Delta_0\cos 2\theta  - b_e &  \Delta_0\sin2\theta\\
 \Delta_0\sin2\theta & \Delta_0\cos 2\theta  + b_e
\end{array}\right)\,,
\eeqa
where we have again discarded a multiple of the identity.
Two comments are in order. Although we are considering a homogeneous
medium, in some of the previous expressions we have retained the
dependence in  $\vec\kappa$ to make more obvious their generalization
to the anisotropic situation. In addition, for the case
of mixing between two Dirac neutrinos, the matrix $U$ is real
(see \Eq{Udos}), but in the case of Majorana neutrinos
three Dirac neutrinos, this is not not necessarily true;
for that reason, in \Eq{overlineHamflavor}
we have not replaced $U^\ast$ by $U$.

Neglecting the right-handed components then, the complete effective
neutrino field is
\beq{eq.r22}
\Psi(x) = \Psi^{(\nu)}_L(x) + \Psi^{(\overline\nu)}_L(x)\,.
\eeq
The wave functions
used to define $\Psi(x)$ must include appropriate
normalization factors $N_{\kappa}$. They  can be calculated in terms
of the self-energy\cite{jn1} and,
in general,  differ from the normalizations in the vacuum.
With the coefficients as given in
\Eq{coefficients}, $N_{\kappa} = 1$  and therefore,
if we take the creation and annihilation operators normalized
such that
\beq{eq.r23}
\{a_{\alpha'}^\ast(\vec\kappa^\prime),a_{\alpha}(\vec\kappa)\} = (2\pi)^3
\delta^{(3)}
(\vec\kappa^\prime - \vec\kappa) \delta_{\alpha^\prime \alpha}\,,
\eeq
then the spin wave functions $\phi_\lambda$ have to be chosen to
satisfy
\beq{normalization}
\phi^\dagger\phi = 1\,.
\eeq

So far we have assumed that the background is
homogeneous.
Now suppose that the matter density
is not constant. Then along the neutrino path
the densities of the different  particles in the  background  become functions
of the time $t$.
The crucial step that we  take is to postulate that, in such situation,
\Eqs{eq.r8}{eq.r9} remain valid and determine
the spinor  wave functions of mixed
left-handed neutrinos with momentum $\vec \kappa$.
Consequently, \Eq{eq.r15} is assumed to give the
associated effective field also
for an inhomogeneous medium.
Of course, in that case \Eq{eq.r10} no longer holds,
and an appropriate solution of \Eq{eq.r9} must be found.
Identical considerations applies to the (right-handed) antineutrinos.
Our \Eq{eq.r9}, with the (time-dependent) Hamiltonian expressed
in terms of $b_e(t) = \sqrt{2} G_Fn_e(t)$,  is  precisely
the Wolfenstein equation that determines the flavor neutrino
evolution through ordinary matter \cite{Wolfenstein}.

\section{Neutrino background}\label{sec:nubackground}

As illustrated in the previous section,
our formalism reproduces the standard
formulas that have been utilized in
the numerous studies of the MSW effect
when it is applied to a medium made of
electrons and nucleons.
In what follows we show how to extend
this field-theoretic approach to take into account
the new features that appear
when several species of neutrinos are also part of the background.
In this situation, the neutrino self-energy receives additional
contributions from the diagrams depicted in Fig.~1,
which involve neutrinos propagating in the
internal lines of the loop.  These contributions are given by
\beqa{eq.r25}
-i\Sigma^{(\nu\nu)}_{ij}(k) & = & \left(-i\frac{g}{2\cos\theta_W}\right)^2\int
\frac{d^4k^\prime}{(2\pi)^4}
\gamma^\mu Li\Delta^{(Z)}_{\mu\nu}(k^\prime - k)
iS_F^{(\nu)}(k^\prime)_{ij}\gamma^\nu L\,,\nonumber\\
& &\\
\label{eq.r25b}
-i\Sigma^{\prime(\nu\nu)}_{ij}(k) & =  &
\delta_{ij}\left(-i\frac{g}{2\cos\theta_W}\right)^2
\gamma^\mu L \left(\frac{ig_{\mu\nu}}{m_Z^2}\right)\nonumber\\
& & \mbox{}\times(-1)\mbox{Tr}\int
\frac{d^4k^\prime}{(2\pi)^4}\left(\sum_l
iS_F^{(\nu)}(k^\prime)_{ll}\right)\gamma^\nu L\,,
\eeqa
where the trace is taken over the Dirac indices.
$S_F^{(\nu)}(k^\prime)$ is the thermal neutrino propagator defined by
\beq{eq.r26}
i\int\frac{d^4k}{(2\pi)^4}e^{-ik\cdot(x - y)}S_F(k)_{ij} =
\langle T\Psi_i(x)\overline\Psi_j(y)\rangle\,,
\eeq
where $T$ stands for the
time-ordered product of the fields.
The angle brackets in \Eq{eq.r26} indicate a
thermal average, defined by
\beq{defaverage}
\langle{\cal O}\rangle = \frac{1}{Z}\sum_i\langle i|\rho{\cal O}|i\rangle\,,
\eeq
for any operator ${\cal O}$,
where
\beq{rho}
\rho = e^{-\beta{\cal H} + \sum_A\alpha_A Q_A}\,,
\eeq
and
\beq{zeta}
Z = \sum_i\langle i|\rho|i\rangle\,.
\eeq
In \Eq{rho} $\beta$ is the inverse temperature, ${\cal H}$ is the Hamiltonian
of the system,
%
%
the $Q_A$ are the (conserved) charges
that commute with ${\cal H}$, and $\alpha_A$ are the chemical
potentials that characterize the composition of the background.

In the ordinary perturbative applications
of FTFT, the right-hand side of \Eq{eq.r26}
is evaluated in terms of the free fields.
Then, a straightforward calculation of the thermal averages
of the products of  creation  and annihilation operators
yields the canonical formulas for $S_F(k)$
in terms of the particle number densities.  In particular, the
propagators for the electron and the nucleons calculated that way,
were used to compute the
contributions to the neutrino self-energy
given in \Eq{coefficients}.
However, for the problem at hand
we cannot use
the thermal free neutrino
propagator to calculate the
additional contributions to the self-energy introduced
by the neutrino mutual interactions.
The reason is that the
neutrino densities are modulated by the oscillation
amplitudes, which evolve according to the equations
we have obtained (Eqs.~(\ref{eq.r9}) and (\ref{eq.r18})),
and the thermal free neutrino
propagator does not incorporate this effect.

Our proposal in this work is
to use, instead of the thermal propagator of the neutrino free-field,
the propagator that corresponds to the effective neutrino
field defined in \Eq{eq.r22}.  The calculation of this propagator proceeds
as in the free field case, simply by substituting
the expansions in Eqs.~(\ref{eq.r15}) and (\ref{eq.r20}) into (\ref{eq.r26}).
However, as will be seen shortly,
for our purpose here, it is not necessary to compute it
explicitly.  The result of a straightforward calculation, whose details
we give below, are
\beqa{eq.r27}
\Sigma^{(\nu\nu)}(k)_{ij} & = & \sqrt{2}G_F(\gamma^\mu L)(j_\mu^{(\nu)} +
j_\mu^{(\overline\nu)})_{ij}\,,\\
\label{eq.r27b}
\Sigma^{\prime(\nu\nu)}(k)_{ij} & = & \delta_{ij}\sqrt{2}G_F(\gamma^\mu L)
\int\frac{d^3\kappa^\prime}{(2\pi)^3}n^\prime_\mu\left[ f_{\nu_e} -
f_{\overline\nu_e} +
(e\rightarrow\mu)\right]\,,
\eeqa
where
\beqa{eq.a6}
j_\mu^{(\nu)}  & = & \int\frac{d^3\kappa^\prime}{(2\pi)^3}n^\prime_\mu\left[
\chi^{(e)} \chi^{(e)\dagger} f_{\nu_e}
+ (e\rightarrow\mu) \right]\,,\nonumber\\
j_\mu^{(\overline\nu)}  & = &
-\int\frac{d^3\kappa^\prime}{(2\pi)^3}n^\prime_\mu\left[ (\chi^{(\overline e)}
\chi^{(\overline e)\dagger})^\ast
f_{\overline \nu_e}
+ (e\rightarrow\mu)\right]\,.
\eeqa
The vector $n^{\prime\mu}$ has the components
\beq{eq.r29}
n^{\prime\mu} = (1,\hat\kappa^\prime)\,,
\eeq
in the rest frame of the medium, and
$f_{\nu_{e,\mu}}$  ($f_{\overline\nu_{e,\mu}}$)
are the momentum distributions, normalized such that
\beqa{eq.a4a}
n^0_{\nu_{e,\mu}}& = & \int
\frac{d^3\kappa^\prime}{(2\pi)^3}f_{\nu_{e,\mu}}\,,\nonumber\\
n^0_{\overline\nu_{e,\mu}} & = & \int
\frac{d^3\kappa^\prime}{(2\pi)^3}f_{\overline\nu_{e,\mu}}\,,
\eeqa
are the number densities  of the flavor neutrinos (antineutrinos) at the
initial time.

Let us  first  derive the result  for
$\Sigma^{(\nu\nu)}$.
Using the
local approximation for the $Z$-boson propagator,
\beq{eq.a1}
\Delta^{(Z)}_{\mu\nu} \simeq \frac{g_{\mu\nu}}{m_Z^2}\,,
\eeq
and the Fierz-like identity
\bdm
\gamma^\alpha LA\gamma_\alpha L = -(\mbox{Tr}A\gamma_\alpha L)\gamma^\alpha
L\,,
\edm
which is valid for any $4\times 4$ matrix $A$, we have
\beq{eq.a2}
\Sigma^{(\nu\nu)}_{ij}(k) = -i\left(\frac{g^2}{4m_W^2}\right)(\gamma^\mu L)
\int\frac{d^4k^\prime}{(2\pi)^4}\mbox{Tr}(S_F(k^\prime)_{ij}\gamma^\mu L)\,.
\eeq
As in \Eq{eq.r25b},  the trace is with respect the Dirac indices.  While we can
calculate the propagator using the definition in \Eq{eq.r26} and then
substitute
the result in the above formula, it is more expedient to observe
that
\beq{eq.a3}
-i\int\frac{d^4k^\prime}{(2\pi)^4}\mbox{Tr}(S_F(k^\prime)_{ij}\gamma^\mu L)
= \langle N\overline\Psi_j(x)\gamma_\mu\Psi_i(x)\rangle\,,
\eeq
which results immediately  from \Eq{eq.r26},
by taking the limit $x\rightarrow y$.
In the last equation $N$ stands for the normal-ordered product.
The quantity in the right-hand side is easily computed by inserting the
plane wave expansions given in Eqs.~(\ref{eq.r15}) and (\ref{eq.r19}).
The statistical
averages are evaluated by means of the formulas
\beqa{eq.a4}
\langle a^\ast_{\alpha^\prime}(\vec\kappa^\prime)a_\alpha(\vec\kappa)\rangle
&= & (2\pi)^3 \delta^{(3)}(\vec\kappa^\prime -
\vec\kappa)\delta_{\alpha^\prime\alpha}f_{\nu_\alpha}\,, \nonumber\\
\langle
a^\ast_{\overline\alpha^\prime}(\vec\kappa^\prime)a_{\overline\alpha}(\vec\kappa)\rangle
&= & (2\pi)^3 \delta^{(3)}(\vec\kappa^\prime -
\vec\kappa)\delta_{\alpha^\prime\alpha}f_{\overline\nu_{\alpha}}\,
\eeqa
where $\alpha = e,\mu$.
In this manner
we obtain
\beq{eq.a5}
\langle N\overline\Psi_j(x)\gamma_\mu\Psi_i(x)\rangle = (j^{(\nu)}_{\mu} +
j^{(\overline\nu)}_{\mu})_{ji}\,,
\eeq
and using Eqs.~(\ref{eq.a5})  and (\ref{eq.a3}) in (\ref {eq.a2})
we finally arrive at \Eq{eq.r27}.
For $\Sigma^{\prime(\nu\nu)}$, by a similar procedure it follows that
\beq{eq.a7}
\Sigma^{\prime(\nu\nu)}(k)_{ij} = \delta_{ij}\frac{g^2}{4m_W^2}(\gamma^\mu
L)\sum_l
 (j^{(\nu)}_{\mu} + j^{(\overline\nu)}_{\mu})_{ll}\,.
\eeq
Since the hamiltonian matrix is hermitian,
\beq{eq.a8}
\chi^{(e,\mu)\dagger}\chi^{(e,\mu)} = 1\,,
\eeq
and
\beq{eq.a9}
\sum_l  (j^{(\nu)}_{\mu} + j^{(\overline\nu)}_{\mu})_{ll} =
\int\frac{d^3\kappa^\prime}{(2\pi)^3}n^\prime_\mu\left[
f_{\nu_e} - f_{\overline\nu_ e} + (e\rightarrow\mu)\right]\,,
\eeq
which gives  \Eq{eq.r27b}.

$\Sigma^{(\nu\nu)}$ can be decomposed as in \Eq{Sigma},
\beq{eq.r31}
\Sigma^{(\nu\nu)} = (a^{(\nu\nu)}\aslash k + b^{(\nu\nu)}\aslash u)L\,,
\eeq
with
\beqa{eq.r32}
a^{(\nu\nu)} & = & \sqrt{2}G_F \int\frac{d^3\kappa^\prime}{(2\pi)^3}
\left(\frac{\hat\kappa^\prime\cdot\hat\kappa}{\kappa}\right)\nonumber\\
&&\mbox{}\times\left[\chi^{(e)} \chi^{(e)\dagger} f_{\nu_e}
-  (\chi^{(\overline e)}
\chi^{(\overline e)\dagger})^\ast
f_{\overline\nu_e} + (e\rightarrow\mu) \right]\,,\nonumber\\
\nonumber\\
b^{(\nu\nu)} & = & \sqrt{2}G_F\int\frac{d^3\kappa^\prime}{(2\pi)^3}
\left(1 - \frac{\hat\kappa^\prime\cdot\hat\kappa}{\kappa}\
\omega\right)\nonumber\\
&&\mbox{}\times\left[\chi^{(e)} \chi^{(e)\dagger} f_{\nu_e}
-  (\chi^{(\overline e)}
\chi^{(\overline e)\dagger})^\ast
f_{\overline\nu_e}\right.
+ \left.(e\rightarrow\mu) \right]\,.
\eeqa
As remarked below \Eq{disprelsol},
the coefficient $a^{(\nu\nu)}$ has a negligible effect
on the dispersion relation and therefore we do not consider it further.
{}From \Eq{eq.r32} and the results of \Ref{jn1},
it can also be verified that the wave function
normalization to which we alluded before
remains equal to one. On the contrary,
$b^{(\nu\nu)}$ produces a new significant
contribution that modifies the hamiltonian $H$
given in Section~\ref{sec:normalmatter}.
Since, for relativistic neutrinos (antineutrinos),
$\omega\approx\kappa$ ($\omega\approx -\kappa$)
the above formula
for $b^{(\nu\nu)}$ translates into the following additional term to the
matrices exhibited in Eqs.~(\ref{Ham}) and (\ref{overlineHam}):
\beqa{eq.r33}
H^{(\nu\nu)} & = & b^{(\nu\nu)}(\kappa,\vec\kappa)\,, \nonumber\\
%
\overline H^{(\nu\nu)} & = & -b^{(\nu\nu)*}
(-\kappa,-\vec\kappa)  =   -H^{(\nu\nu)\ast}\,.
\eeqa
The additional terms produced by $\Sigma^{\prime(\nu\nu)}$
can be obtained in similar fashion; they are proportional
to the identity matrix $I$ and are given by
\beq{eq.r34}
H^{\prime(\nu\nu)} = -\overline H^{\prime(\nu\nu)} =
\sqrt{2}G_FQ_Z^{(\nu\nu)}I\,,
\eeq
where
\beq{eq.r35}
Q_Z^{(\nu\nu)} = \int\frac{d^3\kappa^\prime}{(2\pi)^3}
\left(1 - \hat\kappa^\prime\cdot\hat\kappa\right)\left[f_{\nu_e} -
f_{\overline\nu_e} + (e\rightarrow\mu)\right]\,.
\eeq

To summarize, the propagation of mixed neutrinos through a
background that includes neutrinos
is described by the effective neutrino field $\Psi (x)$ given
in \Eq{eq.r22}.
The time evolution of the amplitudes $\chi^{(\alpha)} (t)$
and $\chi^{(\overline\alpha)} (t)$ that  appear
in the  Fourier expansion of the  field is governed by
\beqa{eq.r36}
i\frac{d\chi^{(e,\mu)}}{dt} = (H^{(matter)} + H^{(\nu\nu)} +
H^{\prime(\nu\nu)})\chi^{(e,\mu)}\,,
\nonumber\\
\nonumber\\
i\frac{d\chi^{(\overline e,\overline\mu)}}{dt} = (\overline H^{(matter)} +
\overline H^{(\nu\nu)} +
\overline H^{\prime(\nu\nu)})\chi^{(\overline e,\overline\mu)}\,,
\eeqa
where $H^{(matter)}$ and $\overline H^{(matter)}$ stand for
the normal-matter
contributions determined in the previous section (Eqs.~(\ref{Ham})
and (\ref{overlineHam})),
with the label $matter$ added to single them out.
Discarding a term proportional to the identity matrix, in the flavor basis
the above equations can be put in the form
\beqa{eq.r41}
i\frac{d\chi^{(e,\mu)}}{dt} & = &
\frac{1}{2}\left(\begin{array}{cc}
-\Delta_0\cos 2\theta  + b_e  + 2h_{ee}&  \Delta_0\sin2\theta + 2h_{e\mu}\\
 \Delta_0\sin2\theta + 2h_{\mu e}& \Delta_0\cos 2\theta  - b_e + 2h_{\mu\mu}
\end{array}\right)\chi^{(e,\mu)}\,,
\nonumber\\
i\frac{d\chi^{(\overline e,\overline\mu)}}{dt} & = &
\frac{1}{2}\left(\begin{array}{cc}
-\Delta_0\cos 2\theta  - b_e  - 2h_{ee}&  \Delta_0\sin2\theta - 2h_{e\mu}\\
 \Delta_0\sin2\theta - 2h_{\mu e}& \Delta_0\cos 2\theta  + b_e - 2h_{\mu\mu}
\end{array}\right)\chi^{(\overline e,\overline\mu)}\,,\nonumber\\
& &
\eeqa
where $h_{\alpha\alpha^\prime}$ are the elements of
\bdm
h\equiv H^{(\nu\nu)} - \frac{1}{2}\mbox{Tr}H^{(\nu\nu)}\,,
\edm
and are explicitly given by
\beqa{eq.r42}
h_{ee} = -h_{\mu\mu}
& = & \sqrt{2}G_F
\int\frac{d^3\kappa^\prime}{(2\pi)^3}
\left(1 - \hat\kappa^\prime\cdot\hat\kappa\right)\nonumber\\
& &\mbox{}\times\frac{1}{2}\sum_{\alpha = e,\mu}\left[\left
(|\chi^{(\alpha)}_e|^2 - |\chi^{(\alpha)}_{\mu}|^2\right)f_{\nu_\alpha}
- \left(|\chi^{(\overline\alpha)}_e|^2 - |\chi^{(\overline\alpha)}_{\mu}|^2
\right)f_{\overline\nu_\alpha}\right]\,,\nonumber\\
\\
\label{eq.r42b}
h_{e\mu}  =  h_{\mu e}^\ast
& = & \sqrt{2}G_F
\int\frac{d^3\kappa^\prime}{(2\pi)^3}
\left(1 - \hat\kappa^\prime\cdot\hat\kappa\right)\nonumber\\
& &\mbox{}\times\sum_{\alpha = e,\mu}\left[
\left(\chi^{(\alpha)}_e\chi^{(\alpha)\ast}_\mu\right)f_{\nu_\alpha}
- \left(\chi^{(\overline\alpha)}_e\chi^{(\overline\alpha)\ast}_\mu\right)
f_{\overline\nu_\alpha}\right]\,.
\eeqa
In the present case, in which we are neglecting
absorptive effects,
we can use the probability conservation conditions
$|\chi^{(\alpha)}|^2 = 1$  and $|\chi^{(\overline\alpha)}|^2 = 1$
in \Eq{eq.r42} and get
\beqa{eq.r42c}
h_{ee} = -h_{\mu\mu} & = & \sqrt{2}G_F
\int\frac{d^3\kappa^\prime}{(2\pi)^3}
\left(1 - \hat\kappa^\prime\cdot\hat\kappa\right)\nonumber\\
& &\mbox{}\times\sum_{\alpha = e,\mu}\left[\left(|\chi^{(\alpha)}_e|^2 -
\frac{1}{2}\right)f_{\nu_\alpha}
- \left(|\chi^{(\overline\alpha)}_e|^2 -
\frac{1}{2}\right)f_{\overline\nu_\alpha}\right]\,.
\eeqa

The set of (non-linear) equations in (\ref{eq.r41}) (or (\ref{eq.r36})),
subject to the initial conditions
given in \Eqs{eq.r13}{eq.r21}, are the ones that must be solved
in the application to an actual physical problem\cite{footnote:fuller}.
They represent the generalization of the Wolfenstein equation to
the physical situation under consideration. In our notation,
$h_{e,\mu}$ and $h_{\mu,e}$ account for the nondiagonal
contributions to the potential energy in the flavor basis,
that arise from the $\nu\nu$ interactions\cite{pantaleone}.

In deriving the previous results no assumption has been
made on the characteristics of the neutrino background.
If the medium is isotropic, then the quantities between brackets
in, for example, \Eq{eq.r32} can not depend on the angular
integration variables.
In such a case, the integral involving the
factor $\hat\kappa^\prime\cdot\hat\kappa$
vanishes and the coefficients in the self-energy decomposition
reduce to
\beqa{eq.r32isot}
a^{(\nu\nu)} & = &  0\,, \nonumber\\
b^{(\nu\nu)} & = & \sqrt{2}G_F\int\frac{d^3\kappa^\prime}{(2\pi)^3}
\left[\chi^{(e)} \chi^{(e)\dagger} f_{\nu_e}
-  (\chi^{(\overline e)}
\chi^{(\overline e)\dagger})^\ast
f_{\overline\nu_e} + (e\rightarrow\mu) \right]\,,\nonumber\\
& &
\eeqa
with similar simplifications in Eqs.~(\ref{eq.r42}) and
(\ref{eq.r42b}).

Usually, the quantity
of interest in applications
is the density of electron neutrinos at a given time.
In our language it is simply
\beq{eq.r38}
n_{\nu_e} = \langle\Psi_{Le}^{(\nu)\dagger}(x)\Psi_{Le}^{(\nu)}(x)\rangle\,,
\eeq
which, upon substituting the expansion given in \Eq{eq.r15} and evaluating
the statistical averages by means of \Eq{eq.a4}, yields
\beq{eq.r39}
n_{\nu_e} = \int\frac{d^3\kappa^\prime}{(2\pi)^3}\left[|\chi_e^{(e)}|^2
f_{\nu_e} + |\chi_e^{(\mu)}|^2 f_{\nu_\mu}\right]\,.
\eeq
The density of neutrinos of other types can be computed in identical manner
in terms of the initial distributions and the flavor amplitudes. For example,
\beqa{eq.r40}
n_{\nu_\mu} & = &
\langle\Psi_{L\mu}^{(\nu)\dagger}(x)\Psi_{L\mu}^{(\nu)}(x)\rangle\nonumber\\
& = & \int\frac{d^3\kappa^\prime}{(2\pi)^3}\left[|\chi_\mu^{(e)}|^2 f_{\nu_e} +
|\chi_\mu^{(\mu)}|^2 f_{\nu_\mu}\right]\,,
\eeqa
with analogous expressions for antineutrinos.  At $t = 0$, these formulas for
$n_{\nu_e}$ and $n_{\nu_\mu}$ reduce to their appropriate
initial values defined in \Eq{eq.a4a}.

At each time $t$, the Hamiltonian matrix in \Eq{eq.r41}
can be diagonalized by the unitary transformation $U_m(t)$
\beq{Umatter}
U_m (t) =
\left(\begin{array}{cc}
e^{i\varphi_m/2}& 0\\
0&e^{-i\varphi_m/2}
\end{array}\right)
\left(\begin{array}{cc}
\cos\theta_m& \sin\theta_m\\
-\sin\theta_m &\cos\theta_m
\end{array}\right)\,,
\eeq
with the time-dependent mixing angles $\theta_m$ and $\varphi_m$
defined by
\beq{eq.r43}
\sin2\theta_m =\frac{\Delta_0\sin2\theta +
2h_{e\mu}}{\sqrt{\left(\Delta_0\cos 2\theta - b_e
- 2h_{ee}\right )^2
+ (\Delta_0\sin 2\theta + 2h_{e\mu})^2}}\,,
\eeq
\beq{eq.r43b}
\varphi_m =  {\rm arg} [\Delta_0\sin2\theta + 2h_{e\mu}]\,.
\eeq
Under the present conditions, the difference between the
(instantaneous) energy eigenvalues is given by
\beq{omegapmprime}
\omega_2 - \omega_1 =
\sqrt{\left(\Delta_0\cos 2\theta  - b_e - 2h_{ee} \right )^2
+ (\Delta_0\sin 2\theta + 2h_{e\mu})^2 }\,.
\eeq

As is evident from \Eq{eq.r43} the neutrino mixing angle
in matter is modified by their
mutual interactions,
and the resonance condition ($\sin2\theta_m = 1$)  becomes
\beq{eq.r44}
\Delta_0\cos 2\theta =  b_e + 2h_{ee}\,.
\eeq
Moreover, the off-diagonal elements of the Hamiltonian
are  in general  complex, and the phase $\varphi_m$ will be
present in contrast  with the ordinary-matter situation.
Similarly,  a unitary  matrix $\overline U_m(t)$ can be introduced
for antineutrinos;  the  respective  formulas for the
angles $\overline\theta_m$ and
$\overline\varphi_m$ are obtained
by replacing  $b_e$, $h_{ee}$, and $h_{e\mu}$ by their
negatives in Eqs.~(\ref{eq.r43}) and (\ref{eq.r43b}).

\section{Absorptive effects}\label{sec:abseffects}

In the previous sections we have
assumed that the corrections of order $g^2/m_W^4$ to the
real part of the self-energy are not important and in the same
spirit we have discarded the imaginary part
of the self-energy.
This is normally justified since the
corrections of order $g^2/m_W^2$, which are proportional to the
particle-antiparticle asymmetries,
give the dominant contributions.
However, this is not true for a CP-symmetric plasma
like the early Universe,
and in such situations the real and the imaginary
part of the neutrino potential become
comparable~\cite{footnote4}.
Our purpose now is to
complement the formalism by examining the
effects induced when the imaginary
part of $\Sigma_{eff}$
is not negligible.

The
damping effects, which are due to the incoherent
interactions of the neutrinos with the particles in the
background, are described
through an anti-hermitian part of the Hamiltonian
that governs the evolution of the flavor amplitudes.
In our treatment,
the damping terms in the Hamiltonian
are determined from the imaginary
part of $\Sigma_{eff}$
in a way that we now explain.

We have learned that the complete Hamiltonian
for neutrinos is
\beq{eq4.1}
H = \kappa + \frac{m^2}{2\kappa}
+ b(\kappa,\vec\kappa)\,,
\eeq
where  $b(\omega,\vec\kappa)$
stands for the coefficient of $\aslash uL$ in the
decomposition of $\Sigma_{eff}$ in \Eq{Sigma}.
In terms of the vector $n^\mu$ defined in
\Eq{eq.r29}, we have
\beqa{eq4.2}
b(\kappa,\vec\kappa)& = &\frac{1}{2}\left(\left.
\mbox{Tr\,}L\aslash n\Sigma_{eff}\right)\right|_{\omega
= \kappa} \nonumber\\
&= &\frac{1}{2\kappa}\overline u_L^0
\Sigma_{eff}(\kappa,\vec\kappa)u_L^0\,,
\eeqa
where, in the second equality, we have
introduced the massless vacuum Dirac spinors $u_L^0$
normalized such that
\beq{eq4.2b}
u_L^0\overline u_L^0 = \kappa L\aslash n\,.
\eeq
If $\Sigma_{eff}$ develops an absorptive part, so that
\beq{eq4.3}
\gamma^0\Sigma_{eff}^\dagger\gamma^0 \not= \Sigma_{eff}\,,
\eeq
the matrix $b$ develops a non-hermitian term.
Thus, if we write
\beqa{eq4.4a}
\Sigma_{eff} & = & \Sigma_r + i\Sigma_{i}\\
\label{eq4.4b}
b & = & b_r + i b_i
\eeqa
where $b_{r,i}$ are hermitian matrices and
\beqa{eq4.5}
\Sigma_r & = & \frac{1}{2}(\Sigma_{eff} +
\gamma^0\Sigma_{eff}^\dagger\gamma^0)\nonumber\\
\Sigma_i & = & \frac{1}{2i}(\Sigma_{eff} -
\gamma^0\Sigma_{eff}^\dagger\gamma^0)
\eeqa
are the dispersive and absorptive part of the
self-energy, then
\beqa{eq4.6}
b_i & = & \frac{1}{2}\left(\left.
\mbox{Tr}\,L\aslash n\Sigma_{i}\right)\right|_{\omega
= \kappa}\nonumber\\
& = & \frac{1}{2\kappa}\overline u_L^0
\Sigma_{i}(\kappa,\vec\kappa)u_L^0\,.
\eeqa
This formula allows us to calculate
the absorptive terms in the evolution equation
for the amplitudes $\chi^{(\alpha)}$, from an explicit
calculation of the absorptive part of the self-energy.
However, using a very general argument,
we will now show that the matrix  $b_i$ is
simply related to the total rates for decay
and inverse decays of the neutrinos.

The starting point of our discussion is the
formal definition
of the elements of the self-energy matrix in the real-time
formulation of FTFT
\beqa{defsigma}
i\Sigma_{21}(z - y)_{\alpha\beta,ij} & = & -\langle\eta_{\alpha,i}(z)
\overline\eta_{\beta,j}(y)\rangle\,,\nonumber\\
i\Sigma_{12}(z - y)_{\alpha\beta,ij} & = & \langle\overline\eta_{\beta,j}(y)
\eta_{\alpha,i}(z)\rangle\,,\nonumber\\
-\Sigma_{11}(z - y) & = & \Sigma_{21}(z - y)\theta(z^0 - y^0)
+ \Sigma_{12}(z - y)\theta(y^0 - z^0)\,,\nonumber\\
-\Sigma_{22}(z - y) & = & \Sigma_{21}(z - y)\theta(y^0 - z^0)
+ \Sigma_{12}(z - y)\theta(z^0 - y^0)\,,
\eeqa
where $\alpha,\beta$ are Dirac indices and $i,j$ are
indices labeling the neutrino mass eigenfields.
$\eta_i$ and $\overline\eta_i$ are the neutrino source
fields, and in terms of them the interaction Lagrangian
is
\beq{Lint}
L_{\mbox{int}} = \overline\nu_L\eta + \overline\eta\nu_L\,.
\eeq
In terms of the elements $\Sigma_{ab}$ the physical
self-energy is given by\cite{coulcov}
\beq{eq4.8}
\Sigma_{eff} = \Sigma_{11} + \Sigma_{12}\,,
\eeq
which from \Eq{defsigma} is immediately recognized to be the
retarded self-energy.  Decomposing $\Sigma_{11}$
in analogy to Eqs.~(\ref{eq4.4a}) and (\ref{eq4.5}),
\Eq{eq4.8} is equivalent to
\beqa{eq4.9}
\Sigma_r(k) & = & \Sigma_{11r}\,,\nonumber\\
\Sigma_i(k) & = & \Sigma_{11i} - i\Sigma_{12}(k)\,.
\eeqa
On the other hand, using the integral representation
of the step function, it follows from \Eq{defsigma}
\beq{eq4.10}
\Sigma_{11i}(k) = \frac{i}{2}\left[\Sigma_{21}(k) +
\Sigma_{12}(k)\right]\,,
\eeq
which implies that $\Sigma_i$, determined from \Eq{eq4.9},
can be equivalently computed using
\beq{eq4.11}
\Sigma_{i}(k) = \frac{i}{2}\left[\Sigma_{21}(k) -
\Sigma_{12}(k)\right]\,.
\eeq
Then from \Eq{eq4.6}
\beq{eq4.6prime}
b_i = \frac{i}{4\kappa}\overline u_L^0\left[\Sigma_{21}(k) -
\Sigma_{12}(k)\right]u_L^0\,,
\eeq
which is a useful formula because the matrix
elements of $\Sigma_{21}$ and
$\Sigma_{12}$ are related to the rates for emission
and absorption of the neutrino.
To see this, we
insert a complete set of states between the field
operators $\eta$ and $\overline\eta$ in \Eq{defsigma}, giving
\beqa{Sigma21specrep}
i\Sigma_{21}(z - y)_{\alpha\beta} & = &
-\frac{1}{Z}\sum_{n,m}e^{-i(q_m - q_n)\cdot
(z - y)}\times\nonumber\\
& & \langle n|\eta(0)|m\rangle\langle m|\overline\eta(0)
|n\rangle Z_n\,,
\eeqa
where $Z$ is defined in \Eq{zeta},
\begin{eqnarray}
Z_n  & = & \langle n|\rho|n\rangle\nonumber\\
& = & e^{-\beta q_n\cdot u + \alpha_n}\,,
\eeqa
and $\alpha_n$ is the eigenvalue of the operator
$\hat{\alpha} = \sum_A\alpha_A Q_A$ corresponding to
the state $|n\rangle$; i.e., $\hat{\alpha}|n\rangle =
\alpha_n|n\rangle$.  From \Eq{Sigma21specrep}
we immediately obtain
\beq{eq4.12}
i\overline u_L^0\Sigma_{21}(\kappa,\vec\kappa)u_L^0
= -2\kappa\Gamma_D
\eeq
where
\beq{eq4.13a}
\Gamma_D = \frac{1}{Z}\frac{1}{2\kappa}\sum_{n,m}
\langle n|\overline u_L^0\eta(0)|m\rangle
\langle m|\overline\eta(0)u_L^0|n\rangle(2\pi)^4
\delta^{(4)}(k + q_n - q_m)Z_n\,.
\eeq
In particular, the diagonal elements $(\Gamma_D)_{ii}$ are the
total rates for the processes $n + \nu_i\rightarrow m$,
averaged over all possible initial
states.  The off-diagonal
elements have a similar structure but they involve
the product of the amplitudes for two different neutrino species.
In similar fashion,
\beq{eq4.13b}
i\overline u_L^0\Sigma_{12}(\kappa,\vec\kappa)u_L^0
= 2\kappa\Gamma_I
\eeq
where $\Gamma_I$ is the matrix
\beq{GammaInv}
\Gamma_I = \frac{1}{Z}\frac{1}{2\kappa}\sum_{n,m}
\langle n|\overline u_L^0\eta(0)|m\rangle
\langle m|\overline\eta(0)u_L^0|n\rangle(2\pi)^4
\delta^{(4)}(k + q_n - q_m)Z_m\,,
\eeq
whose diagonal elements are the total rates
for the inverse processes $m\rightarrow n + \nu_i$,
averaged over the initial states.
Then using the results of Eqs.~(\ref{eq4.12}) and (\ref{eq4.13b})
in \Eq{eq4.6prime}
we finally arrive at
\beq{eq4.15}
b_i = -\frac{\Gamma}{2} \,,
\eeq
with
\beq{eq4.15b}
\Gamma = \Gamma_D + \Gamma_I \,.
\eeq

In situations of exact equilibrium,
the $\nu_{Li}$ share the same chemical potential
$\alpha_\nu$.  In that case,
since $\eta$ has
the same quantum numbers as the neutrino field,
then
\bdm
\langle n|\overline u(p)\eta(0)|m\rangle \not= 0
\edm
only for those states such that
$\alpha_m = \alpha_n + \alpha_\nu\,$.
Thus, in \Eq{GammaInv} we can replace
\begin{equation}
Z_m = e^{-\beta\kappa + \alpha_\nu}Z_n\,,
\eeq
which yields
%
\beq{eq4.16}
\Gamma = (1 + e^{-x})\Gamma_D\,,
\eeq
with $x = \beta\kappa - \alpha_\nu$.
However in situations that do
not correspond to exact equilibrium the formula
in \Eq{eq4.15b} is the appropriate one.

Then in summary, if we denote by $H_r$ the energy matrix
displayed in \Eq{eq.r36}
\beq{eq4.17}
H_r = H^{(matter)} + H^{(\nu\nu)} + H^{\prime(\nu\nu)}\,,
\eeq
then the complete Hamiltonian describing the attenuation
effects induced by the incoherent neutrino interactions is
\beq{eq4.18}
H = H_r - i\frac{\Gamma}{2}\,.
\eeq
We remark that, since the total
probability is not conserved any more, the replacements
\bdm
|\chi^{(\alpha,\overline\alpha)}_\mu|^2 =
1 - |\chi^{(\alpha,\overline\alpha)}_e|^2
\edm
made in \Eq{eq.r42c}
are not valid, and the correct formulas
for $h_{ee}$ and $h_{\mu\mu}$ are those given in \Eq{eq.r42}
in this case.

For the antineutrinos, the expression equivalent to
\Eq{eq4.1} is
\beq{eq4.1anti}
\overline H =
\kappa + \frac{m^2}{2\kappa}
- b^\ast(-\kappa,-\vec\kappa)\,,
\eeq
which determines the evolution
of the  amplitudes $\chi^{(\overline\alpha)}$.
Since
\beqa{eq4.2anti}
b^\ast(-\kappa,-\vec\kappa) & = &
\frac{1}{2}\left(
\mbox{Tr\,}L\aslash n\Sigma_{eff}(-\kappa,-\vec\kappa)\right)\nonumber\\
& = & \frac{1}{2\kappa}\overline u_L^0
\Sigma_{eff}(-\kappa,-\vec\kappa)u_L^0\,,
\eeqa
it follows that the complete Hamiltonian becomes
\beq{eq4.18anti}
\overline H =
\overline H_r - i\frac{\overline\Gamma}{2}\,,
\eeq
where
\beqa{eq4.17anti}
\overline H_r &= & \overline H^{(matter)} + \overline H^{(\nu\nu)} +
\overline H^{\prime(\nu\nu)}\,, \nonumber\\
\overline\Gamma & = &\overline\Gamma_D + \overline\Gamma_I\,.
\eeqa
$\overline\Gamma_{D,I}$ are given by the same formulas
for $\Gamma_{D,I}$ in Eqs.~(\ref{eq4.13a}) and (\ref{GammaInv}),
but with
the substitution $k\rightarrow -k$ in the delta function.
By crossing then,
it follows that the diagonal
elements of $\overline\Gamma_D$
and $\overline\Gamma_I$ are the rates for the processes
$n\rightarrow m + \overline\nu_i$ and
$m + \overline\nu_i\rightarrow n$, respectively\cite{footnote:vspinors}.

There is one final comment we wish to make.
The precise identification that we have made
between the anti-hermitian part of the
Hamiltonian and the total rate for decays
and inverse decays is intuitively appealing and
useful for practical purposes.  While ordinarily
that allows us to determine the anti-hermitian
part of the Hamiltonian by computing the relevant
rates, there is one situation
in which it does not.  If the background contains
neutrinos, the rates for processes involving the background
neutrinos depends on their number densities which,
as already remarked, are modulated by the oscillation
mechanism.  It is then not obvious how to
calculate the rates for such processes
by the traditional way of calculating transition
probability amplitudes.  However, as a byproduct
of our formalism we obtain the following very precise
procedure.
We calculate the $21$ and $12$ elements
of the self-energy matrix using the FTFT Feynman rules
and then from Eqs.~(\ref{eq4.13a})
and (\ref{GammaInv}) obtain $\Gamma_{D,I}$. The dependence
on the neutrino number densities shows up in the
diagrams in which background neutrinos propagate in the
internal lines of the loops.  Thus, by using in those
diagrams the same
effective neutrino propagator that was introduced
in Section~\ref{sec:nubackground} to calculate
the effects of the neutrino background on the
dispersive terms,
the effect of the modulation of the
number densities by the oscillating amplitudes is
properly taken into account in this case also.

\newpage
\begin{center}{\bf Figure Captions}
\end{center}
\begin{description}
\item[Fig. 1.]
Neutrino background contribution to the self-energy.

\end{description}


\newpage
\begin{thebibliography}{99}

\bibitem{FTFT} A comprehensive review of the general
	formalism, with an extensive list
			of references, is given by N. P. Landsman and Ch. G. van Weert,
			\physrep{145}{141}{1987}.  For the application to neutrinos see,
			D. Notzold and G. Raffelt, \npb{307}{924}{1988};
   P. B. Pal and T. N. Pham, \prd{40}{714}{1989};
			J. F. Nieves, \prd{40}{866}{1989}.
   J. F. Nieves in {\em Particles and Fields}, IV Mexican
    School, edited by J. L. Lucio and A. Zepeda (World Scientific,
	Singapore
   1992).

\bibitem{Wolfenstein} L. Wolfenstein, \prd{17}{2369}{1978},
   {\bf 20}, 2634 (1979);
    V. Barger, K. Whisnant, S. Pakvasa and R. J. N. Phillips,
   	\ibid{22}{2718}{1980};
  	 P.~Langacker, J.~Leveille and J.~Sheiman, \ibid{27}{1228}{1983}.

\bibitem{MSmirnov}S.~P.~Mikheyev and A.~Yu.~Smirnov,
	Yad. Phyz. {\bf 42} (1985) 1441
    [Sov. J. Nucl. Phys. {\bf 42} (1985) 913]; Nuovo Cimento
    {\bf C9} (1986) 17;
    H. A. Bethe, \prl{56}{1305}{1986}.

\bibitem{earlierlit} 	P. Langacker, S. T. Petcov,
	G. Steigman and S. Toshev,
   \npb{282}{589}{1987};
         K. Kainulainen, \plb{244}{191}{1990};
         K. Enqvist, K. Kainulainen and J. Maalampi,
	\plb{249}{531}{1990};
	K. Enqvist, J. Maalampi,  J. T. Pieltoniemi,
	\npb{358}{435}{1991};
 	R. Barbieri and A. Dolgov, \plb{237}{440}{1990};
	M. J Thompson and B. J. McKellar, \plb{259}{113}{1991};
		G.M. Fuller, R.W. Mayle, J.R. Wilson
		and D.N. Schramm, \astropj{322}{795}{1987};
	J. C. D'Olivo and Manuel Torres, in {\em Relativity
   and Gravitation: Classical and Quantum}, Proceedings
   of the SILARG VII, Cocoyoc, Mexico 1990,
   edited by J. C. D'Olivo et. al. (World Scientific, Singapore, 1991),
   p. 356.  M. J. Savage, R. A. Malaney and G. M. Fuller,
	\astropj{368}{1}{1991};
   G. M. Fuller, R. Mayle, B. S. Meyer, and J. R. Wilson,
	\astropj{389}{517}{1992}.

\bibitem{cline} J. M. Cline, \prl{68}{3137}{1992}.

\bibitem{densitymatrix}
	K. Enqvist, K. Kainulainen and J. Maalampi,
	\npb{349}{754}{1991};
	R. Barbieri and A. Dolgov, \npb{349}{743}{1991};
	K. Enqvist, K. Kainulainen,  M. Thomson,
	\npb{373}{498}{1992};
	X. Shi, D. N. Schramm, and B. D. Fields,
	\prd{48}{2563}{1993}.


\bibitem{pantaleone} J. Pantaleone,
	\plb{287}{128}{1992}; \prd{46}{510}{1992}.
         See also, J. C. D'Olivo and J. F. Nieves,
	\npb{35}{466}{1994} (Proc. Suppl.).

\bibitem{samuel} S. Samuel, \prd{48}{1462}{1993};
          V. A. Kosteleck\'{y} and S. Samuel,
	 \ibid{49}{1740}{1994}.

\bibitem{mswreviews} For reviews of neutrino oscillations in matter see,
			S.~P.~Mikheyev and A.~Yu.~Smirnov, Usp. Fiz. Nauk. {\bf 153} (1987) 3,
			[Sov. Phys. Usp.{\bf 30} (1987) 759 ];
			T.~K.~Kuo and J.~Pantaleone, \rmp{61}{937}{1989};
			P. B. Pal, Int. J. Mod. Phys. {\bf A7} (1992) 5387.

\bibitem{solarnus} J. N. Bahcall, {\em Neutrino Astrophysics}
				(Cambridge University Press, Cambridge, England, 1989).
For a review of the present status of the solar
neutrino problem,  see   N. Hata  University of  Pennsylvania Report
No. UPR-0612T, 1994 .


\bibitem{densitymatrix2}  A. D. Dolgov,
	\sovphys{33}{700}{1981};
				L. Stodolsky, \prd{36}{2273}{1987}.
\bibitem{siglraffelt} G. Sigl and G. Raffelt,
	\npb{406}{423}{1993}.

\bibitem{mckellar} B. H. J. McKellar and M. J. Thomson,
	\prd{49}{2710}{1994}.

\bibitem{fuller} Yong-Zhong Qian and G. M. Fuller, University
of Washington preprint ``Neutrino-Neutrino
Scattering and Matter-Enhanced Neutrino Flavor Transformation in
Supernovae'', 1994.

\bibitem{pantaleone2} J. Pantaleone,
	Indiana University preprint IUHET-276, revised May 1994.

\bibitem{Weldon:fermions} H. A. Weldon, \prd{26}{2789}{1982}.
   As pointed in this reference,
   terms of the form $\sigma_{\mu\nu}k^\mu u^\nu$ do not appear
   at the one-loop level and for this reason
   we exclude them from the beginning.

\bibitem{DNT} The results can be easily inferred from
  	 the formulas given in D'Olivo, Nieves and Torres,
		\prd{46}{1172}{1992}, or  D. Notzold and G. Raffelt in \Ref{FTFT}.
	We are assuming that the corrections of order $g^2/m_W^4$ to the
 	real part of the self-energy are not important. In the same
   	spirit we are discarding the imaginary part
  	of the self-energy, which implies that
   	the absorptive effects on the propagation of the neutrino
   	can be neglected.  This last assumption will be
	relaxed in Section~\ref{sec:abseffects},
	where we consider the damping effects
	induced by the imaginary part of the self-energy.

\bibitem{jn1} J. F. Nieves, \prd{40}{866}{1989}.
	As shown there,
	$N_\kappa = (\partial f/\partial\omega)|_{\omega = \kappa}$,
	where $f = (1 - a)(\omega - \kappa) - b$.

\bibitem{footnote:fuller} The formulas given in \Eq{eq.r36} are equivalent
to the ones of \Ref{fuller},  which were written
by adapting the general (density-matrix) approach of \Ref{siglraffelt}
to the problem of neutrino propagation and
transformations in a supernova.

\bibitem{footnote4}  Within the framework of the FTFT,
	the real contributions of order
	$g^2/m_W^4$ arise from the momentum-dependent terms of
	the boson propagators in the neutrino self-energy
	diagrams~\cite{DNT}.
	In order to calculate them
	for a neutrino
	background, we have to go back to \Eq{eq.r25} and
	keep the next terms
	in the momentum expansion of the $Z$-boson propagator.

\bibitem{coulcov} A careful discussion of the
justification and implications of
this definition is given by J. C. D'Olivo
and J. F. Nieves, ``Coulomb and Covariant Gauges
in Finite Temperature QED'', preprint LTP-043-UPR,
July 1994.

\bibitem{footnote:vspinors} Notice that for massless
	particles the vacuum spinors $u_L^0$ and $v_L^0$
	coincide.

\end{thebibliography}
\end{document}